\documentclass[superscriptaddress,reprint,showpacs,aps,pra,longbibliography]{revtex4-1}
\usepackage{graphics}
\usepackage{bm}
\usepackage{bbm}
\usepackage{amsmath}
\usepackage{amssymb}
\usepackage{graphicx}
\usepackage{amsthm}

\begin{document}

\title{Many-body entanglement  in fermion systems}
\author{N.\ Gigena}\altaffiliation[Present address: ]{Faculty of Physics, University of Warsaw, Pasteura 5, 02-093 Warsaw, Poland}
\affiliation{IFLP/CONICET and Departamento  de F\'{\i}sica,
    Universidad Nacional de La Plata, C.C. 67, La Plata (1900), Argentina}
\author{M.\ Di Tullio}
\affiliation{IFLP/CONICET and Departamento  de F\'{\i}sica,
    Universidad Nacional de La Plata, C.C. 67, La Plata (1900), Argentina}
\author{R.\ Rossignoli}
\affiliation{IFLP/CONICET and Departamento  de F\'{\i}sica,
    Universidad Nacional de La Plata, C.C. 67, La Plata (1900), Argentina}
\affiliation{Comisi\'on de Investigaciones Cient\'{\i}ficas (CIC), La Plata (1900), Argentina}

\begin{abstract}
We discuss a general bipartite-like representation and Schmidt decomposition of an arbitrary  pure state of $N$ indistinguishable fermions, based on states of $M<N$ and $(N$-$M)$ fermions.  It is directly connected with the reduced $M$- and $(N$-$M)$-body density matrices (DMs), which have the same spectrum  in such states. The concept of $M$-body entanglement  emerges  naturally in this scenario, generalizing that of one-body entanglement. Rigorous majorization relations satisfied by the normalized $M$-body DM are then  derived, which imply that the associated  entropy  will not increase, on average, under a class of  operations which have these DMs as post-measurement states. Moreover, such entropy is an upper bound to  the bipartite entanglement entropy generated by a class of operations which map the original state to a bipartite state of $M$ and $N-M$ effectively distinguishable fermions.  Analytic evaluation of the spectrum of  $M$-body DMs in some strongly correlated fermionic states is also provided. 
\end{abstract}
\maketitle

\section{introduction}

A remarkable feature of quantum mechanics is the existence of correlations between quantum systems that cannot be emulated by their classical counterpart. Entanglement is the most celebrated manifestation of such correlations and it has been object of intense research in quantum physics,  particularly within the field of quantum information theory \cite{NC.00}. Particle indistinguishability is another fundamental feature of quantum mechanics, lying at the heart of condensed matter physics and quantum field theories. An interesting problem combining these two fundamental concepts is that of the study and quantification of correlations between indistinguishable particles, a topic that has received increasing attention in the last years \cite{BFFM.20}. Indistinguishability poses a nontrivial difficulty in the study of quantum correlations, because the notion of entanglement is intimately connected with that of local operations, and the latter are possible only if the constituents of the system can be distinguished. Different approaches to this problem have been considered, like mode entanglement \cite{Za.02,Shi.03,FL.13}, extensions based on correlations between observables \cite{BK.04, ZL.04,SI.11, BG.13,BF.14}
and entanglement beyond symmetrization
\cite{SC.01,SL.01,ES.02,GM.02,PY.01,WV.03,IV.13,OK.13,SL.141,GR.15,MB.16}, which is independent of the choice of single particle (sp) basis. The relation between these forms of entanglement has been analysed by different authors
\cite{BFFM.20,FL.13,WV.03,GR.15,DD.16,GR.17,BF.17,DGR.18,DRGC.19,SD.19,DS.20,DI.20} as well as the question regarding whether  symmetrization  correlations should be addressed as entanglement \cite{CM.07,KC.14,CC.18,FC.18,MY.19}. 

In this paper we explore the  generalization of the bipartite formulation devised in \cite{GDR.20} for the notion of one-body entanglement, a measure of correlations beyond (anti)symmetrization introduced in \cite{GR.15}. We start by considering  operators on the fermion Fock space of the system creating general  $M$-fermion states, which are used to show that a general $N$-fermion state can be always  written as a bipartite-like state of $M<N$ and $N-M$ fermions. This bipartite representation is connected with the definition of the $M$ and $N-M$-body reduced density matrices (DMs) \cite{CK.61,An.63,AF.17} in a way that closely resembles the case of bipartite states of distinguishable constituents. The ensuing $(M,N-M)$ Schmidt-like decomposition of the state determines the diagonal form of these DMs,  entailing they share the same non-zero eigenvalues. Pushing forward this analogy we propose to link the correlations between $M$ and $N-M$-body observables, which we call {\it $M$-body entanglement}, to the mixedness of the $M$-body DM, as formalized by eigenvalue majorization. We introduce a family of entropic measures of such correlations and show that there exists a class of operations not increasing the amount of these correlations in any  $N$-fermion state. Finally, we prove the existence of a family of quantum maps converting $N$-fermion states into bipartite states of effectively distinguishable $M$ and $N-M$-fermions, in the sense of occupying orthogonal sp  subspaces.  Conversion by means of any of these maps is such that the entanglement entropy of the bipartite target state is bounded from above by the $M$-body  entropy, assigning the latter a clear operational meaning.  Explicit examples of $M$-body DMs and their eigenvalues in some physical states are also provided. 

\section{Formalism}

We consider a sp space $\cal H$ of finite dimension $D$, spanned by fermion operators $c_i$, $c^\dag_i$, $i=1,\ldots,D$ satisfying the anticommutation relations 
\begin{equation}\{c_i,c^\dag_j\}=\delta_{ij},\;
\{c_i,c_j\}=\{c^\dag_i,c^\dag_j\}=0\,.
\label{cnm}
\end{equation}
We also define the $M$-fermion creation  operators
\begin{equation}
    C^{(M)\dag}_{\bm{\alpha}}=c^\dag_{i_1}\ldots c^\dag_{i_M}\,,\label{Cm} 
\end{equation}
where $i_1<i_2<\ldots<i_M$ and $\bm{\alpha}=(i_1,\ldots,i_M)$  labels all $\binom{D}{M}=\frac{D!}{M!(D-M)!}$ distinct sets of $M$  sp states sorted in increasing order.  These operators satisfy    \begin{eqnarray}
    \langle 0|C^{(M)}_{\bm{\alpha}} C^{(M')\dag}_{\bm{\alpha'}}|0\rangle&=&\delta^{MM'}\delta_{\bm{\alpha\alpha'}}\,,\label{nor}\\
    \sum_{\bm{\alpha}} C^{(M)\dag}_{\bm{\alpha}} 
     C^{(M)}_{\bm{\alpha}}&=&\binom{\hat{N}}{M}\,,\label{Nm}
\end{eqnarray}
where $\hat{N}=\sum_i c^\dag_i c_i$ is the fermion number operator and $\binom{\hat{N}}{M}$ is the operator taking the value $\binom{N}{M}$ in a state of $N$ fermions ($\binom{\hat{N}}{M}|\Psi\rangle=\binom{N}{M}|\Psi\rangle$ for $\hat{N}|\Psi\rangle=N|\Psi\rangle$, with $\binom{\hat{N}}{1}=\hat{N}$, $\binom{\hat{N}}{2}=\frac{\hat{N}^2-\hat{N}}{2}$, etc.).   Eq.\ \eqref{Nm} is a generalization of the number operator,  representing the   number of  ``$M$-fermion composites''.  
The states $C^{M\dag}_{\bm{\alpha}}|0\rangle$  are Slater Determinants (SDs) and  form, for all $\bm{\alpha}$ and $0\leq M\leq D$,   an orthonormal basis of the $2^D$-dimensional Fock space associated to ${\cal H}$.  

A normalized pure state $|\Psi\rangle$ of $N$ fermions ($\hat{N}|\Psi\rangle=N|\Psi\rangle$) can  then be expanded in this basis as 
\begin{eqnarray}
    |\Psi\rangle&=&\frac{1}{N!}\sum_{i_1,\ldots, i_N} \!\!\Gamma_{i_1\ldots i_N}\, c^\dagger_{i_1}\ldots  c^\dagger_{i_N}|0\rangle\label{psi1}\\
    &=&\sum_{\bm{\alpha}}\Gamma^{(N)}_{\bm{\alpha}} C^{(N)\dag}_{\bm{\alpha}}|0\rangle\,,\label{psin}
\end{eqnarray}
where $\Gamma_{i_1\ldots i_N}$ are the elements of a fully antisymmetric tensor and  $\Gamma^{(N)}_{\bm{\alpha}}=\langle 0|C^{(N)}_{\bm{\alpha}}|\Psi\rangle=
\Gamma_{i_1\ldots i_N}$ (for  $\bm{\alpha}=(i_1,\ldots,i_n)$, $i_1<i_2<\ldots<i_N$), 
with 
\begin{equation}\sum_{\bm{\alpha}} |\Gamma^{(N)}_{\bm{\alpha}}|^2=
\frac{1}{N!}\sum_{i_1,\ldots, i_N}|\Gamma_{i_1\ldots i_N}|^2=1\,.\label{Norm}\end{equation}
Thus, $|\Gamma^{(N)}_{\bm{\alpha}}|^2=\langle\Psi|C_{\bm\alpha}^{(M)\dag}C_{\bm\alpha}^{(M)}|\Psi\rangle$ is the probability of finding the $N$ sp states  ${\bm{\alpha}}$ occupied in $|\Psi\rangle$.     
\subsection{The $(M,N-M)$   representation and the \\$M$-body DM}
For $0\leq M\leq N$ we can also  rewrite the state  \eqref{psi1} as 
   \begin{equation}
       |\Psi\rangle=
       {\textstyle\binom{N}{M}^{-1}}\sum_{\bm{\alpha},\bm{\beta}} \Gamma^{(M)}_{\bm{\alpha}\bm{\beta}}C^{(M)\dag}_{\bm\alpha} C^{(N-M)\dag}_{\bm\beta}|0\rangle\,,
       \label{Psm}
   \end{equation}
where $\Gamma^{(M)}_{\bm{\alpha\beta}}\equiv\Gamma^{(M,N-M)}_{\bm{\alpha\beta}}$ is given by  \begin{equation}
\Gamma^{(M)}_{\bm{\alpha}\bm{\beta}}=\langle 0|C^{(N-M)}_{\bm{\beta}} C^{(M)}_{\bm{\alpha}}|\Psi\rangle=
\Gamma_{i_1\ldots i_M j_{1}\ldots j_{N-M}}\,,\label{gdef}
\end{equation} 
 for $\bm{\alpha}=(i_1,\ldots,i_M)$, 
 $\bm{\beta}=(j_1,\ldots,j_{N-M})$,
 and the sum in \eqref{Psm} is over all  $\binom{D}{M}$ and $\binom{D}{N-M}$ distinct sets of $M$ and $N-M$ sp states respectively, with  \eqref{Norm} implying 
 \begin{equation}\sum_{\bm{\alpha},
\bm{\beta}}\left|\Gamma^{(M)}_{\bm{\alpha\beta}}\right|^2
=\binom{N}{M}\,.\label{norm}
\end{equation}
 We will denote the expression \eqref{Psm} as the {\it $(M,N-M)$-body  
 representation} of the $N$-fermion state $\eqref{psi1}$. It is a  bipartite-like  expansion of $|\Psi\rangle$ in orthogonal $M$- and $(N-M)$-fermion states,  leading to a  $\binom{D}{M}\times \binom{D}{N-M}$ matrix representation $\Gamma^{(M)}$ of the original tensor $\Gamma$ in \eqref{psi1}. Of course, decompositions $(M,N-M)$ and $(N-M,M)$ are equivalent, with   \begin{equation}\Gamma^{(N-M)}=
  (-1)^{M(N-M)}(\Gamma^{(M)})^T\,,\label{trs}
  \end{equation} 
 due the antisymmetry of $\Gamma$ ($T$ denotes transpose). Eq.\ \eqref{psin} is the trivial $(N,0)$ representation.  

  From the antisymmetry of $\Gamma$ it also follows that 
  \begin{eqnarray}C^{(M)}_{\bm\alpha}|\Psi\rangle&=&
    \sum_{\bm\beta} \Gamma^{(M)}_{\bm{\alpha\beta}}C^{(N-M)\dag }_{\bm{\beta}}|0\rangle\,,
    \label{3}
    \end{eqnarray}
which represents the (unnormalized) state of the remaining $N-M$ fermions when the $M$ sp states labelled by $\bm\alpha$ are occupied in $|\Psi\rangle$. 
  Eqs.\ \eqref{3} and  \eqref{nor} imply that  the  {\it 
 $M$-body density matrix} \cite{CK.61,AF.17},  whose elements   are  defined as 
    \begin{eqnarray}\rho^{(M)}_{
\bm{\alpha\alpha'}}&:=&\langle\Psi|C^{(M)\dag}_{\bm{\alpha'}}C^{(M)}_{\bm\alpha}|\Psi\rangle\,,\label{M}
    \end{eqnarray}
    can be expressed in terms of $\Gamma^{(M)}$ as 
    \begin{equation}
\rho^{(M)}=\Gamma^{(M)}\Gamma^{(M)\dag}\,,\label{mDM}
\end{equation}
i.e.\ $\rho^{(M)}_{\bm{\alpha\alpha'}}=\sum_\beta \Gamma^{(M)}_{\bm{\alpha\beta}}\Gamma^{(M)\,*}_{\bm{\alpha'\beta}}$,  in the same way as the  reduced density matrix $\rho^A={\rm Tr}_B|\Psi_{AB}\rangle\langle\Psi_{AB}|$ is   obtained from a general state $|\Psi_{AB}\rangle=\sum_{i,j} G_{ij}|i_A,j_B
\rangle$ of a system of two distinguishable components ($\rho^{A}_{ii'}=(GG^\dag)_{ii'}$ for $\rho^A_{ii'}=\langle\Psi_{AB}||i'_A\rangle\langle i_A|\otimes \mathbbm{1}_B|\Psi_{AB}\rangle$). 
 In particular, 
$\rho^{(M)}_{\bm{\alpha\alpha}}$ is the probability of finding the $M$ sp states specified by  $\bm{\alpha}$ occupied in $|\Psi\rangle$. Eq.\ \eqref{mDM}  is a positive semidefinite $\binom{D}{M}\times\binom{D}{M}$ matrix which determines the average of any $M$-body operator $\hat{O}^{(M)}=\sum_{\bm{\alpha},\bm{\alpha'}}O^{(M)}_{\bm{\alpha\alpha}'}C^{(M)\dag}_{\bm\alpha}C^{(M)}_{\bm{\alpha}'}$ through 
\begin{equation}
\langle\Psi|\hat{O}^{(M)}|\Psi\rangle=
 {\rm Tr}\,[\rho^{(M)}O^{(M)}]\,.\label{Om}\end{equation}
Its trace is given by 
\begin{equation}
    {\rm Tr}\,[\rho^{(M)}]=\binom{N}{M}\,,\label{trm}
\end{equation}
as implied by \eqref{Nm} or \eqref{norm}. We also notice that 
    \begin{equation}C^{(N-M)}_{\beta}|\Psi\rangle=\sum_{\alpha}\Gamma^{(N-M)}_{\beta\alpha}C_\alpha^{(M)\dag}|0\rangle\,.\end{equation} 
    Hence, using \eqref{trs} the partner $(N-M)$-body DM, of elements     $\rho^{(N-M)}_{\beta\beta'}=\langle\Psi|C^{(N-M)\dag}_{\beta'}C^{(N-M)}_\beta|\Psi\rangle$,  is  
           \begin{equation}
    \rho^{(N-M)}=\Gamma^{(N-M)}\Gamma^{(N-M)\dag}=(\Gamma^{(M)})^T\Gamma^{(M)*}\,,\label{pDMs}
    \end{equation}
which  shows it has {\it the same non-zero eigenvalues} (and hence the same trace) as the $M$-body DM \eqref{mDM} \cite{CK.61,AF.17}, as discussed in \ref{IIB}  in more detail. 
 
In particular, for $M=1$, $C^{(1)\dag}_{\bm\alpha}=c^\dag_i$ and we recover from \eqref{Psm} the $(1,N-1)$ representation \cite{GDR.20}
 \begin{equation} |\Psi\rangle=\frac{1}{N}\sum_{i,\bm\alpha} \Gamma^{(1)}_{i\bm\alpha}c^\dag_i C^{(N-1)\dag}_{\bm\alpha}|0\rangle\,,\label{Ps1}\end{equation}
where the $D\times \binom{D}{N-1}$ matrix $\Gamma^{(1)}$  determines the {\it one-body} DM (also denoted as SPDM)  $\rho^{(1)}_{ii'}=\langle\Psi|c^\dag _{i'}c_{i}|\Psi\rangle$ through 
 \begin{equation} \rho^{(1)}=\Gamma^{(1)}\Gamma^{(1)^\dag}\label{rh1}\,.\end{equation}
 
  We finally remark that Eq.\ \eqref{Psm} is a particular case of the more general {\it $k$-partite representation}   of the state \eqref{psi1},
 \begin{eqnarray}|\Psi\rangle&=&{\textstyle\frac{M_1!\ldots M_k!}{N!}}\!\!
 \sum_{\bm\alpha_1,\ldots,\bm\alpha_k}\!\!\!\!\Gamma^{M_1\ldots M_k}_{\bm\alpha_1\ldots\bm\alpha_k}C^{(M_1)\dag}_{\bm\alpha_1}\ldots C^{(M_k)\dag}_{\bm\alpha_k}|0\rangle\,,\;\;\;
  \end{eqnarray}
 where 
$\Gamma^{M_1\ldots M_k}_{\bm\alpha_1\ldots\bm\alpha_k}=
\langle 0|C^{(M_k)}_{\bm\alpha_k}\ldots C^{(M_1)}_{\bm\alpha_1}|\Psi\rangle=\Gamma_{i_1\ldots i_N}$  for $\bm\alpha_1\equiv(i_1,\ldots,i_{M_1}), \ldots,\bm\alpha_k\equiv(i_{N-M_k+1},\ldots,i_N)$ 
and $\sum_{j=1}^k M_j=N$, with  $k\leq N$ and  sums running over all $\binom{D}{M_j}$ sets of $M_j$ sp states.  The basic expansion \eqref{psi1} corresponds to $k=N$ and $M_j=1$ for $j=1,\ldots,N$.  

\subsection{The $(M,N-M)$ Schmidt representation
     \label{IIB}}
We can now employ the  singular value decomposition of the matrix  $\Gamma^{(M)}$, 
\begin{subequations}
\begin{eqnarray}
    \Gamma^{(M)}&=&U^{(M)}D^{(M)}V^{(N-M)\dag}\,,\\\ D^{(M)}_{\nu\nu'}&=&\sqrt{\lambda^{(M)}_\nu}\delta_{\nu\nu'}\,,\end{eqnarray} \label{SVD}
    \end{subequations}
$\!\!$where  $U^{(M)},V^{(N-M)}$ are $\binom{D}{M}\times\binom{D}{M}$ and    $\binom{D}{N-M}\times\binom{D}{N-M}$ unitary matrices 
 and $D^{(M)}\equiv D^{(M,N-M)}$ a $\binom{D}{M}\times \binom{D}{N-M}$ diagonal matrix with non-negative elements. Here  $\lambda^{(M)}_\nu$ denote the square of the singular values of $\Gamma^{(M)}$, i.e.\ the   eigenvalues of $\Gamma^{(M)}\Gamma^{(M)\dag}=\rho^{(M)}$ or equivalently 
$\Gamma^{(M)T}\Gamma^{(M)*}=\rho^{(N-M)}$, which have the same spectrum (except for the number of zero eigenvalues). It then becomes  possible to rewrite Eq.\ \eqref{Psm} in the Schmidt-like diagonal form  
\begin{equation}
    |\Psi\rangle={\textstyle\binom{N}{M}^{-1}}\sum_{\nu=1}^{n_{M}} \sqrt{\lambda^{(M)}_\nu} A^{(M)\dag}_\nu B^{(N-M)\dag}_\nu|0\rangle\,,\label{ScDc}
      \end{equation}
where $n_M$ is the rank of $\Gamma^{(M)}$ and 
\begin{eqnarray}A^{(M)\dag}_\nu&=&\sum_{\bm\alpha}U^{(M)}_{\bm\alpha\nu}
C^{(M)\dag}_{\bm\alpha}\,,\label{A1}\\
B^{(N-M)\,\dag}_\nu&=&\sum_{\bm\beta} V_{\bm\beta\nu}^{(N-M)*}C^{(N-M)\,\dag}_{\bm\beta}\,,\label{A2}\end{eqnarray}
are ``collective'' operators creating, respectively, $M$ and $N-M$ fermions in generally entangled (i.e.,  non-SD's for $M\geq 2$) states. Nevertheless, since they are unitarily related to the original operators $C^{(M)\dag}_{\bm\alpha}$ and $C^{(N-M)\dag}_{\bm\beta}$, 
 they  still satisfy,  for  $1\leq M\leq N-1$,
  \begin{eqnarray}
        \langle 0| A^{(M)}_{\nu} A^{(M)\dag}_{\nu'}|0\rangle&=&\delta_{\nu\nu'}=\langle 0| B^{(N-M)}_{\nu}B^{(N-M)\dag}_{\nu'}|0\rangle
        ,\;\;\;\;\;\;\label{nor2}\\
        \sum_\nu A^{(M)\dag}_\nu A^{(M)}_\nu&=&\binom{\hat{N}}{M}=\sum_\nu B^{(N-M)\dag}_\nu B^{(N-M)}_\nu
        ,  \;\;\;\;\;\;\;    \label{Nm2}\\
        \langle 0|B^{(N-M)}_{\nu'}A^{(M)}_{\nu}|\Psi\rangle&=&\sqrt{\lambda^{(M)}_\nu}\delta_{\nu\nu'}\,.
        \label{Adiag}\end{eqnarray}
 
 Moreover, Eqs.\ \eqref{3}, \eqref{SVD} and \eqref{A1} lead to  
\begin{equation}A^{(M)}_\nu|\Psi\rangle=\sqrt{\lambda_\nu^{(M)}} B^{(N-M)\,\dag}_\nu|0\rangle\,,
\label{Ap}\end{equation}
such that $B^{(N-M)\dag}_\nu|0\rangle$ is the state  of remaining $N-M$ fermions when $M$ fermions are measured to be in the ``normal'' or ``natural'' state $A^{(M)\dag}_\nu|0\rangle$. Eqs.\ \eqref{nor2}--\eqref{Ap}  also imply that the normal operators $A^{(M)}_\nu$, $B^{(N-M)}_\nu$  {\it diagonalize} the $M$- and $(N-M)$-body DM's:  
\begin{eqnarray}\langle\Psi|A^{(M)\dag}_{\nu'}A^{(M)}_{\nu}|\Psi\rangle&=&(U^{(M)\dag}\rho^{(M)}U^{(M)})_{\nu\nu'}\nonumber\\&=&\lambda_\nu^{(M)}\delta_{\nu\nu'}\label{diagm}\\
&=&
\langle\Psi|B^{(N-M)\,\dag}_{\nu'} B^{(N-M)}_{\nu}|\Psi\rangle\,.
\nonumber\end{eqnarray}  
For $M=1$ we recover from \eqref{A1}--\eqref{A2} the diagonal representation of the one-body DM  $\rho^{(1)}$ \cite{GDR.20}, with $A^{(1)\dag}_\nu=\sum_i U_{i\nu}c^\dag_i=c^\dag_\nu$  the operators creating a fermion in the ensuing natural sp orbitals. 

In the trivial case $M=N$, $\rho^{(N)}$ has a single non-zero eigenvalue $\lambda^{(N)}_1=1$ associated with the operator $A^{(N)\,\dag}_1=\sum_{\bm\alpha} \Gamma^{(N)}_{\bm\alpha} C^{(N)\dag}_{\bm\alpha}$ creating the state. On the other hand, in an $N$-fermion SD, which can be always written as 
$|\Psi\rangle=c^\dag_1\ldots c^\dag_N|0\rangle$ by a suitable choice of the  operators $c^\dag_i$,  $\rho^{(M)}$ has  just 
$\binom{N}{M}$ non-zero eigenvalues $\lambda^{(M)}_\nu=1$, associated with 
the $\binom{N}{M}$ operators $A^{(M)\dag}_{\nu}=c^\dag_{i_1}\ldots c^\dag_{i_M}$, $1\leq i_1<\ldots<i_M\leq N$,  with support on the  $N$ occupied sp states, satisfying 
$\langle \Psi|A^{(M)\dag}_\nu A^{(M)}_{\nu'}|\Psi\rangle=\delta_{\nu\nu'}$.   For instance, the  decomposition \eqref{ScDc} of an $N=3$ SD $c^\dag_1 c^\dag_2 c^\dag_3|0\rangle$ for $M=1$ is just  $|\Psi\rangle=\frac{1}{3}\sum_{i=1}^3c^\dag_iB^{(2)\dag}_i|0\rangle$, with  $B^{(2)\dag}_1=c^\dag_2c^\dag_3$, $B^{(2)\dag}_2=-c^\dag_1 c^\dag_3$, 
$B^{(2)\dag}_3=c^\dag_1c^\dag_2$ and 
$\langle c^\dag_ic_j\rangle=\delta_{ij}=\langle B^{(2)\dag}_iB^{(2)}_j\rangle$ for $i,j\leq 3$. Similarly  for general $N$.  Thus, in a SD $\rho^{(M)}$ is idempotent:  $(\rho^{(M)})^2=\rho^{(M)}$ $\forall M\leq N$.

And for a general pure  two-fermion state  $|\Psi_2\rangle=\frac{1}{2}\sum_{i<j}\Gamma_{ij}c^\dag_ic^\dag_j|0\rangle$, with $\Gamma_{ij}=-\Gamma_{ji}$, 
the (non-zero) singular values of $\Gamma^{(1)}=\Gamma$ for the $M=1$ decomposition $(1,1)$,   and hence the eigenvalues of $\rho^{(1)}=\Gamma \Gamma^\dag$, 
are always two-fold degenerate \cite{SC.01}, such that the natural operators can be paired as  $A^{(1)\dag}_\nu=c^\dag_\nu$,  $A^{(1)\dag}_{\bar{\nu}}=c^\dag_{\bar{\nu}}$ 
for $\lambda^{(1)}_\nu=\lambda^{(1)}_{\bar{\nu}}$, with 
$B^{(1)}_{\nu}=c^\dag_{\bar{\nu}}$, $B^{(1)}_{\bar\nu}=-c^\dag_{\nu}$ and 
$c^\dag_{\nu(\bar{\nu})}=\sum_j U_{j\nu(\bar{\nu})}c^\dag_j$. Then Eq.\ \eqref{ScDc} leads to  
\begin{subequations}\begin{eqnarray}
|\Psi_2\rangle&=&\frac{1}{2}
\sum_{\nu}\sqrt{\lambda_\nu^{(1)}}\, (A^{(1)\dag}_\nu B^{(1)\dag}_\nu +A^{(1)\dag}_{\bar\nu}B^{(1)\dag}_{\bar\nu})|0\rangle\;\;\;\;\\
&=&
\sum_\nu\sqrt{\lambda_\nu^{(1)}}\,c^\dag_\nu c^\dag_{\bar\nu}|0\rangle\,,\label{tfs}
\end{eqnarray}
\end{subequations}
where \eqref{tfs} is the well-known Slater decomposition of a two-fermion state \cite{SC.01,ES.02}, 
with $\langle c^\dag_\nu c_{\nu'}\rangle=\langle c^\dag_{\bar\nu} c_{\bar\nu'}\rangle=\lambda^{(1)}_\nu\delta_{\nu\nu'}$, 
 $\langle c^\dag_\nu c_{\bar\nu'}\rangle=0$ and $\sum_\nu\lambda^{(1)}_\nu=1$. 
 For $D=4$ (two fermions in 4 sp levels), 
 there are just two distinct eigenvalues $\lambda_\nu^{(1)}$, given by 
 $\lambda_{\pm}=\frac{1\pm\sqrt{1-C^2}}{2}$, with $C=2|\Gamma_{12}\Gamma_{34}-\Gamma_{13}\Gamma_{24}+\Gamma_{14}\Gamma_{23}|=2\sqrt{\lambda_+\lambda_-}$ the fermionic concurrence \cite{SC.01,ES.02,GR.15}, and \eqref{tfs} can be rewritten as  $(\sqrt{\lambda_+}c^\dag_1 c^\dag_{\bar{1}}+ \sqrt{\lambda_-}c^\dag_2 c^\dag_{\bar{2}})|0\rangle$.  

\subsection{The eigenvalues of the $M$-body DM} 
While the eigenvalues 
$\lambda^{(1)}_\nu=\langle\Psi|c^\dag_\nu c_\nu|\Psi\rangle$ of the SPDM $\rho^{(1)}$ always lie in the interval $[0,1]$ (as $c^\dag_\nu$ are standard fermion  operators),  for $M\geq 2$ 
those of $\rho^{(M)}$  can be {\it greater than 1} when $|\Psi\rangle$ is not a SD, 
since  the  normal operators $A^{(M)\dag}_\nu$ will in general no longer be a single product of $M$ fermion creation operators. These operators  may exhibit  boson-like features for even $M\geq 2$, or in general features of  boson + fermion creation operators for odd $M\geq 3$. 

We first note that any operator  $A^{(M)\dag}=\sum_{\bm\alpha}\gamma_{\bm\alpha}C^{(M)\dag}_{\bm\alpha}$ creating $M<N$ fermions can be expanded  in the normal operators \eqref{A1} as  $A^{(M)\dag}=\sum_\nu \gamma_\nu A^{(M)\dag}_\nu$, with 
$\gamma_\nu=\sum_{\bm\alpha} U_{{\bm\alpha}\nu}^*\gamma_{\bm\alpha}$. 
Hence, using \eqref{diagm}, 
\begin{equation}\langle \Psi|A^{(M)\dag} A^{(M)}|\Psi\rangle=\sum_\nu \lambda^{(M)}_\nu|\gamma_\nu|^2\,, \label{avnu}\end{equation}
with $\langle 0|A^{(M)}A^{(M)\dag}|0\rangle
=\sum_{\nu}|\gamma_\nu|^2$. Thus  for  normalized operators  satisfying $\langle 0|A^{(M)}A^{(M)\dag}|0\rangle=1$, 
the largest eigenvalue $\lambda^{(M)}_{\rm max}$ of $\rho^{(M)}$ bounds {\it any} such average:  
\begin{equation}\langle \Psi|A^{(M)\dag}A^{(M)}|\Psi\rangle\leq \lambda^{(M)}_{\rm max}\,.\label{bnd}\end{equation}
In a SD,  $\lambda^{(M)}_{\rm max}=1$ 
and $\langle \Psi|A^{(M)\dag}A^{(M)}|\Psi\rangle\leq 1$ $\forall$ $M$ and  normalized $A^{(M)\dag}$. This bound can of course  be broken in more general fermion states. 

For instance, let us define, 
assuming even sp dimension $D$,   the collective pair creation operator \begin{equation}
A^{\dag}=\frac{1}{\sqrt{D/2}}
\sum_{i=1}^{D/2}c^{\dag}_{2i-1}c^\dag_{2i}\,,
\label{Ad}
\end{equation}
which satisfies $[A^\dag,\hat{N}]=-2A^\dag$ and 
\begin{equation}
    [A,A^\dag]=1-\tfrac{2}{D}\hat{N}\label{AC}\,,
\end{equation}
implying $\langle 0|AA^\dag|0\rangle=1$. 
We then consider the normalized states ($0\leq k\leq D/2$) 
\begin{eqnarray}
    |\Psi_{2k}\rangle&=&\frac{(A^\dag)^k}{k!\sqrt{(\frac{2}{D})^k\binom{D/2}{k}}}|0\rangle
  = \frac{\sum_\mu C^{(2k)\dag}_\mu}{\sqrt{\binom{D/2}{k}}}|0\rangle\,,
\label{ks}\end{eqnarray}
which contain  $k$ of such pairs and hence  $N=2k$ fermions, and where 
$C^{(2k)\dag}_\mu=\prod_i(c^\dag_{2i-1}c^\dag_{2i})^{n_{i\mu}}$ with $n_{i\mu}=0,1$,  $\sum_{\mu} n_{i\mu}=k$ and $1\leq \mu\leq \binom{D/2}{k}$.  They  satisfy 
\begin{subequations}
\begin{eqnarray}A^\dag|\Psi_{2k}\rangle&=&{\textstyle\sqrt{(k+1)(1-2k/D)}}|\Psi_{2(k+1)}
\rangle\\
    A|\Psi_{2k}\rangle&=&{\textstyle\sqrt{k(1-2\frac{k-1}{D})}}|\Psi_{2(k-1)}\rangle\,.\end{eqnarray}\label{A2k}
\end{subequations}
so that for these states $A^\dag$  behaves as a perfect ladder operator and can therefore be considered an ideal coboson according to the definition given in \cite{L.05}. 

Since in $|\Psi_{2k}\rangle$ all sp states have the same probability of being occupied, and fermions are created in pairs, 
it is apparent that the  SPDM $\rho^{(1)}$ will have  just a single degenerate eigenvalue $\lambda^{(1)}_\nu=N/D \leq 1$ (see Appendix A).  Then it will be  uniformly mixed, i.e.,  proportional to the identity  and  hence diagonal in {\it any} sp basis:  
\begin{equation}
\rho^{(1)}=\frac{2k}{D}\mathbbm{1}\,,\label{r1kd}
\end{equation}
Hence the states \eqref{ks} lead to  {\it maximum one-body entanglement} compatible with a given value of $N$, with 
\eqref{r1kd} showing explicitly that they are not SDs for $2k<D$. 

In contrast,  from Eqs.\ \eqref{A2k} it follows that $A^\dag A|\Psi_{2k}\rangle=k(1-2\frac{k-1}{D})|\Psi_{2k}\rangle$, implying  that the two-body DM $\rho^{(2)}$ has   one  large nondegenerate  eigenvalue (see Appendix A)  
\begin{eqnarray} 
\lambda_{\rm max}^{(2)}&=&\langle\Psi_{2k}| A^\dag A|\Psi_{2k}\rangle
=k\left(1-2\frac{k-1}{D}\right)
\geq 1\,,\label{l21}
\end{eqnarray}
associated with the normal operator $A^\dag$, {\it which satisfies}    
$\lambda_{\rm max}^{(2)}>1$ for $2\leq k\leq \frac{D}{2}-1$ 
[$\lambda^{(2)}_{\rm max}=1+(k-1)(1-\tfrac{2k}{D})$].  All remaining $\binom{D}{2}-1$ eigenvalues  of $\rho^{(2)}$ are small and identical to (see Appendix A) 
\begin{equation}\lambda_{\rm rest}^{(2)}= \frac{4k(k-1)}{D(D-2)}\leq 1 \,,\label{l22}\end{equation} 
 such that $\lambda^{(2)}_{\rm max}+(\binom{D}{2}-1) \lambda_{\rm rest}^{(2)}=\binom{N}{2}$ (Eq.\   \eqref{trm}).  Since $\langle \Psi|A^\dag A|\Psi\rangle\leq 1 $ in {\it any} SD (Eq.\ \eqref{bnd}), \eqref{l21}  clearly signals as well that $|\Psi_{2k}\rangle$ is not a SD. 

The dominant eigenvalue \eqref{l21} is maximum in the half-filled case $k=[\frac{D+2}{4}]$, where $\lambda^{(2)}_{{\rm max}}\approx D(1+2/D)^2/8$ increases linearly with $D$ for large $D$, and can then become arbitrarily large. 
For $D\gg k$,     $\lambda^{(2)}_{\rm max}\approx k=N/2$  is just the number of pairs,  whereas $\lambda^{(2)}_{\rm min}\approx (2k/D)^2$ becomes very small. As seen from \eqref{AC}--\eqref{ks}, 
 for $D\rightarrow\infty$ at fixed $N$,  $A^\dag$ becomes a ``true'' boson 
 ($[A,A^\dag]\rightarrow 1$), with   $|\Psi_{2k}\rangle\rightarrow \frac{(A^\dag)^k}{\sqrt{k!}}|0\rangle$  a condensate of $k$-bosons. 
 
\begin{figure}
    \centering
    \hspace*{-0.5cm}\includegraphics[scale=0.65]{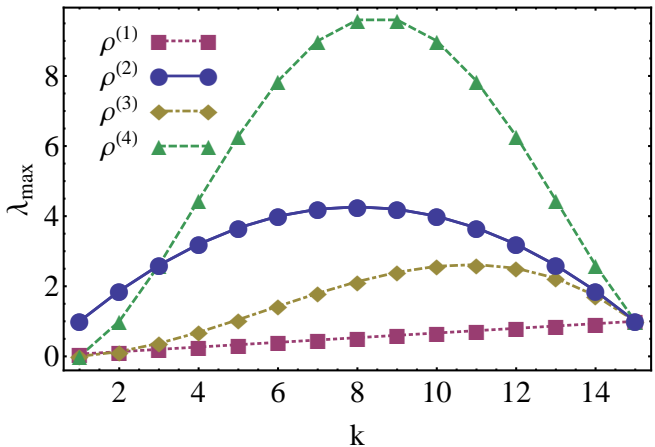}
    \caption{Maximum eigenvalue $\lambda_{\rm max}=\lambda^{(M)}_{\rm max}$ of the $M$-body density matrix $\rho^{(M)}$ for $M\leq 4$ in the state (\ref{ks}), as a function of the number of pairs $k$, for sp space dimension $D=30$ (all labels and quantities plotted are dimensionless).}
    \label{fig1}
\end{figure}

 Eigenvalues of odd $M$  DM's can also exceed  $1$. For instance, in an odd state $|\Psi_{2k+1}\rangle=c^\dag_{D+1}|\Psi_{2k}\rangle$, where we have enlarged the sp space with one additional state, the largest eigenvalue  of the three-body DM $\rho^{(3)}$  is again given by (\ref{l21}): $\langle\Psi_{2k+1}|c^\dag_{D+1}A^\dag A\, c_{D+1}|\Psi_{2k+1}\rangle=
\lambda^{(2)}_{\rm max}\geq 1$. 
It corresponds to the number of pairs  times the number of ``single fermions'' (1).   The eigenvalues of $\rho^{(3)}$ in the even state \eqref{ks} can also be analytically determined (see Appendix A). Its largest  eigenvalue, 
\begin{equation}
    {\textstyle\lambda^{(3)}_{\rm max}=\frac{2k(k-1)(1-\frac{2}{D}(k-1))}{D-2}\,,}\label{l31}
\end{equation}
while smaller than \eqref{l21}, still satisfies $\lambda^{(3)}_{\rm max}>1$ 
 for $1+\sqrt{D/2}<k<D/2$, reaching $\approx \frac{2D}{27}$ 
at  $k\approx \frac{D}{3}$ for  $D\gg 1$. Fig.\ \ref{fig1} depicts  $\lambda^{(M)}_{\rm max}$ vs.\ $k$ for $M\leq 4$ in the states \eqref{ks}. 

From  \eqref{A2k} it also follows that $(A^\dag)^m A^m|\Psi_{2k}\rangle\propto|\Psi_{2k}\rangle$ for $m\leq k$. Thus,  the largest  eigenvalue  of the $2m$-body DM in the state  \eqref{ks} is associated to the  normalized operator  
$A^{(2m)\dag}=\frac{(A^\dag)^m}{m!\sqrt{(\frac{2}{D})^m\binom{D/2}{m}}}$ (see also Appendix A):   
\begin{equation} \lambda^{(2m)}_{\rm max}=\langle \Psi_{2k}|\frac{(A^\dag)^mA^m}{m!^2(\frac{2}{D})^m\binom{D/2}{m}}|\Psi_{2k}\rangle=
\frac{\binom{k}{m}\binom{D/2-k+m}{m}}
{\binom{D/2}{m}}\,,\label{l2m}\end{equation}
which generalizes Eq.\ \eqref{l21} ($m=1$ case). Thus, $\lambda^{(2m)}_{\rm max}>1$ for 
$m<k<D/2$, with $\lambda^{(2m)}_{\rm max}\approx \binom{k}{m}$ if $D\gg k$. 
  Similarly, in the  odd state $|\Psi_{2k+1}\rangle=c^{\dag}_{D+1}|\Psi_{2k}\rangle$, 
$\lambda_{\rm max}^{(2m+1)}=\lambda_{\rm max}^{(2m)}$. 

The eigenvalues of $\rho^{(M)}$ can also be all smaller than $1$. For instance, in a $N=D/2$ GHZ-like fermion state 
\begin{equation}|\Psi_{D/2}\rangle=\frac{1}{\sqrt{2}}(c^\dag_{1}\ldots c^\dag_{D/2}+c^\dag_{D/2+1}\ldots c^\dag_D)|0\rangle\label{GHZ}\,,\end{equation}
all nonzero eigenvalues of $\rho^{(M)}$ are easily seen to be  
\begin{equation}
\lambda^{(M)}_\nu=1/2\,,
\end{equation} 
for $1\leq M\leq N-1$, $2\binom{N}{M}$ degenerate. This example shows that distinct  $N$-fermion states which appear {\it  identical} at the one-body level, 
like \eqref{ks} and  \eqref{GHZ} for $N=D/2$ ($\rho^{(1)}=\frac{1}{2}\mathbbm{1}$), can differ significantly at higher $M$-body  levels ($\lambda^{(2)}_{1}=1/2$ in \eqref{GHZ} while $\lambda^{(2)}_{1}=\frac{D}{8}(1+\frac{4}{D})>1$ in \eqref{ks} for $k=D/4$ and $D\geq 8$). 

\section{$M$-body Entanglement}
\subsection{Mixedness of the $M$-body DM} 
The Schmidt-like decomposition \eqref{ScDc}  of an $N$-fermion pure state, and the fact that the positive numbers $\lambda^{(M)}_\nu$ represent the non-zero eigenvalues of both the $M$ and  $(N-M)$-body DM's, naturally lead to a notion  of  $M$-body entanglement based on the spread of these eigenvalues. It characterizes the correlations between $M$ and $(N-M)$-body observables in an $N$-fermion state. More precisely, given 
 two pure states $|\Psi\rangle$, $|\Phi\rangle$ of $N$ fermions, we will say that $|\Psi\rangle$ is {\it not less $(M,N-M)$ entangled}, or simply {\it not less $M$-body entangled} than $|\Phi\rangle$, if $\rho^{(M)}_\Psi$ is {\it more mixed than} (or equally mixed as)  $\rho^{(M)}_\Phi$, i.e., if their  eigenvalues satisfy the majorization relation 
\begin{equation}
    \bm{\lambda}(\rho_{\Psi}^{(M)})\prec
    \bm{\lambda}({\rho}_{\Phi}^{(M)})\label{Em}\,,
\end{equation}
where $\bm{\lambda}(\rho^{(M)})$  denotes the  spectrum of $\rho^{(M)}$, sorted in decreasing order. Explicitly, Eq.\ \eqref{Em} means that all inequalities \cite{Bh.97,MOA.11}
\begin{equation}\sum_{\nu=1}^j
\lambda_\nu(\rho^{(M)}_\Psi)\leq \sum_{\nu=1}^j\lambda_\nu(\rho^{(M)}_\Phi)
\,,\;\;j=1,\ldots,{\textstyle\binom{D}{M}}-1\,,\label{lam}\end{equation}
are to be satisfied by the sorted eigenvalues $\lambda_\nu$ of $\rho^{(M)}_{\Psi}$ and $\rho^{(M)}_{\Phi}$, 
with equality for $j=\binom{D}{M}$, implying  that those  of $\rho_\Psi^{(M)}$ are more spread out than those of ${\rho}_\Phi^{(M)}$.  Of course, 
one may likewise employ the partner  DM  $\rho^{(N-M)}$ in \eqref{Em}--\eqref{lam}  since they share the same non-zero eigenvalues.  
For $M=1$ we recover the concept of  one-body entanglement, determined by the SPDM $\rho^{(1)}$  \cite{GDR.20,GR.15}. 

Eq.\ \eqref{Em} is analogous to that satisfied by  local reduced states in systems of distinguishable components (where it warrants that $|\Psi\rangle$ can be converted to $|\Phi\rangle$ by LOCC operations \cite{N.99,NC.00,NV.01}).  Notice,  however, that majorization provides a partial order, entailing that two states may be uncomparable according to previous criterion.  

For example, in an $N$-fermion SD,   $\rho^{(M)}$ presents just  $\binom{N}{M}$ non-zero eigenvalues equal to $1$, while in the  GHZ-like state \eqref{GHZ}, all $2\binom{N}{M}$ non-zero eigenvalues of $\rho^{(M)}$ are equal to $1/2$ for $M<N$, implying
\begin{equation} 
\bm{\lambda}(\rho^{(M)}_{\Psi_{D/2}})\prec\bm{\lambda}(\rho^{(M)}_{\rm SD})\end{equation}
for $1\leq M\leq N-1$. Then 
the state \eqref{GHZ} is more entangled than a SD at {\it any} $M$-body level   ($1\leq M\leq N-1$).  

However, in the pair condensate $|\Psi_{2k}\rangle$ of Eq.\ \eqref{ks}  with $2\leq k\leq D/2-1$, 
while clearly $\bm{\lambda}(\rho^{(1)}_{\Psi_{2k}})\prec
\bm{\lambda}(\rho^{(1)}_{\rm SD})$ 
(Eq.\ \eqref{r1kd}), { neither $\bm{\lambda}(\rho^{(2)}_{ \Psi_{2k}})\prec\bm{\lambda}(\rho^{(2)}_{\rm SD})$ nor  $\bm{\lambda}(\rho^{(2)}_{\rm SD})\prec\bm{\lambda}(\rho^{(2)}_{\Psi_{2k}})$ are fulfilled},  since the largest eigenvalue of $\rho^{(2)}_{\Psi_{2k}}$ is greater than $1$ whereas remaining $\binom{D}{2}-1$ eigenvalues are non-zero and lower than $1$ 
(Eqs.\ \eqref{l21}--\eqref{l22}), with  $\binom{D}{2}>\binom{N}{2}$ for $D>N$. 
Hence, SDs no longer provide the least mixed two-body DM. The same occurs with  the three-body DM when its largest eigenvalue  in the state \eqref{ks} exceeds $1$ (Eq.\ \eqref{l31}), in which case $\bm{\lambda}(\rho^{(3)}_{\Psi_{2k}})\nprec\bm{\lambda}(\rho^{(3)}_{\rm SD})$ and $\bm{\lambda}(\rho^{(3)}_{\rm SD})\nprec\bm{\lambda}(\rho^{(3)}_{\Psi_{2k}})$. 

\subsection{$M$-body entropies}
Associated with previous definition 
\eqref{Em}, we may first consider  the  $M$-body entropies   \begin{subequations}   
\begin{eqnarray}
E_f^{(M)}(|\Psi\rangle)&=&S_f(\rho_{\Psi}^{(M)})=S_f(\rho_{\Psi}^{(N-M)})\,,\label{Emf}\\
&=&\sum_\nu f[\lambda_\nu(\rho^{(M)}_\Psi)]\,,
\label{Emfb}
\end{eqnarray} 
\label{Emff}
\end{subequations}
where  $S_f(\rho)={\rm Tr}f(\rho)$ is a trace-form entropy \cite{CR.02,RC.99}, with  $f:\mathbb{R}_{\geq 0}\rightarrow\mathbb{R}_{\geq 0}$
 a strictly concave non-negative function complying with $f(0)=0$. These entropies  will satisfy  
 \begin{equation}
 E_f^{(M)}(|\Psi\rangle)\geq E_f^{(M)}(|\Phi\rangle)\end{equation} whenever Eq.\ \eqref{Em} is  fulfilled  \cite{Bh.97,MOA.11,RC.022}.  
 
Eq.\ \eqref{Emff} is particularly suitable for defining a one-body entanglement entropy \cite{GR.15,GDR.20} since $\lambda^{(1)}_\nu\in[0,1]$ and standard entropic measures  can be employed. For $M\geq 2$ 
it is possible to employ measures such as the bosonic-like  entropy, obtained for  
$f(\lambda)=-\lambda\log\lambda + (1+\lambda)\log(1+\lambda)$ 
(such that $\sum_\nu f(\lambda_\nu)$ represents the von Neumann entropy of independent bosons in the grand canonical ensemble with average occupation numbers $\lambda_\nu$ \cite{RC.99}), which is just an example of an increasing concave function of $\lambda$ satisfying $f(0)=0$.  

A second possibility, strongly  motivated by the   majorization relations  derived in the next subsections, is to consider the entropy of the normalized densities  
\begin{equation}\rho_n^{(M)}=\rho^{(M)}/{\textstyle\binom{N}{M}}
\label{rhon}\,,\end{equation} 
 which have  eigenvalues  $\lambda^{(M)}_\nu/\binom{N}{M}\in[0,1]$ 
 and satisfy  ${\rm Tr}[\rho_n^{(M)}]=1$. In this case we may employ any entropic measure $S_f(\rho)={\rm Tr}\,f(\rho)$ intended for standard probabilities  and define an associated $M$-body  entropy as 
\begin{eqnarray}E_{n_f}^{(M)}(|\Psi\rangle)&=&
S_f(\rho_{n\Psi}^{(M)})=S_f(\rho_{n\Psi}^{(N-M)})
\label{Emfn}\,.\end{eqnarray}
In particular, the von Neumann entropy $S(\rho)=-{\rm Tr}\rho\log_2\rho$ leads to 
 $S(\rho^{(M)}_n)=S(\rho^{(M)})/\binom{N}{M}+\log_2\binom{N}{M}$. 
 Other Schur-concave functions \cite{Bh.97} of $\rho_n^{(M)}$ can also be used. 
 Since at fixed  $N$  Eq.\ \eqref{Em} is fully equivalent to 
\begin{equation}\bm{\lambda}(\rho^{(M)}_{n\Psi})\prec\bm{\lambda}(\rho^{(M)}_{n\Phi})\,,\label{precn}\end{equation} 
  as all eigenvalues are just rescaled by the same factor, it will also imply  $E_{n_f}^{(M)}(|\Psi\rangle)\geq E_{n_f}^{(M)}(|\Phi\rangle)$. And any pair of $N$-fermion states states uncomparable with \eqref{Em} will remain uncomparable with \eqref{precn}, and viceversa.  
 
On the other hand, Eq.\ \eqref{precn} can  also be used to compare the mixedness of reduced DMs $\rho^{(M)}_n$  for states with distinct $N$, as $\rho^{(M)}_n$ has fixed trace, implying 
\begin{equation}
   \bm{\lambda}(\rho_{n\Psi}^{(M)})\prec\bm{\lambda}(\rho_{n\Phi}^{(M)})\Rightarrow E_{n_f}^{(M)}(|\Psi\rangle)\geq E_{n_f}^{(M)}(|\Phi\rangle) \,. \label{ineq2}
\end{equation}
We finally remark that the converse of Eq.\ \eqref{ineq2}  
does not hold in general: Only if the entropic inequality $E_{n_f}^{(M)}(|\Psi\rangle)\geq E_{n_f}^{(M)}(|\Phi\rangle)$ holds for {\it all} concave functions $f$ (and not just a particular choice) it can be ensured  that  $\bm{\lambda}(\rho_{n\Psi}^{(M)})\prec\bm{\lambda}(\rho_{n\Phi}^{(M)})$ \cite{RC.022}. And this implies    $\bm{\lambda}(\rho_{\Psi }^{(M)})\prec\bm{\lambda}(\rho_{\Phi }^{(M)})$ only when $\rho_{\Psi }^{(M)}$ and $\rho_{\Phi}^{(M)}$ have the same trace, i.e.\  $|\Psi\rangle$ and $|\Phi\rangle$  the same fermion number.  
For states with distinct $N$ Eq.\ \eqref{Em} is to be replaced by \eqref{precn}. 

\subsection{Operations not increasing the   $M$-body entropy}
Let us now determine the basic operations which do not  increase the $M$-body entropy \eqref{Emfn}. 
We first note that one-body unitary transformations 
\begin{equation}|\Psi\rangle\rightarrow {\cal U}|\Psi\rangle\,,\;\;{\cal U}=\exp[-i \sum_{i,j}H_{ij}c^\dag_i c_j]\,,\end{equation} 
where  $H^\dag=H$,  lead to unitary transformations of all $M$-body DMs, thus  leaving their eigenvalues  and hence $M$-body entanglement unchanged: 
Since ${\cal U}^\dag c_i {\cal U}=\sum_j U_{ij} c_j$ with $U=\exp[-iH]$, then $\rho^{(1)}\rightarrow U\rho^{(1)} U^\dag$  and $\rho^{(M)}\rightarrow 
U^{(M)}\rho^{(M)}U^{(M)\dag}$, 
with $U^{(M)}_{\bm{\alpha\alpha}'}=\epsilon_{i_1\ldots i_m}U_{\alpha_1\alpha'_{i_1}}\ldots U_{\alpha_m\alpha'_{i_m}}$ and $\epsilon$ the fully antisymmetric tensor.

We now show, for both pure states $|\Psi\rangle$ with fixed fermion number $N$ and also  mixed states  \begin{equation}\rho=\sum_i q_i|\Psi_i\rangle\langle\Psi_i|\,,\label{rhox}\end{equation} 
with definite $N$ ($q_i\geq 0$, $\sum_i q_i=1$ and $\hat{N}|\Psi_i\rangle=N|\Psi_i\rangle$ $\forall$ $i$),  the following theorem: \\
{\it Theorem 1.  
The quantum operation described by the Kraus operators 
\begin{equation}
{\cal M}^{(1)}_j=\frac{c_j}{\sqrt{N}}\,,\;\;j=1,\ldots,D\label{M2}
\end{equation}
which satisfy $\sum_{j=1}^D {{\cal M}^{(1)}_j}^\dag {\cal M}^{(1)}_j=\hat{N}/N=\mathbbm{1}_N$ within the subspace of states  with definite fermion number $N$,   does not increase, on average, the mixedness of the  normalized $M$-body DMs $\rho^{(M)}_n=\rho^{(M)}/\binom{N}{M}$ 
for $1\leq M\leq N-1$, and implies the majorization relation 
\begin{equation}
\bm{\lambda}(\rho^{(M)}_n)\prec \sum_j p_j\bm{\lambda}(\rho^{(M)}_{jn})\,,\label{maj2}\end{equation}
between the spectrum of the initial and post-measurement normalized $M$-body DM's.} 

In \eqref{maj2}  $\rho^{(M)}_{jn}=\rho^{(M)}_j/\binom{N-1}{M}$  are the normalized $M$-body  DMs determined by the post-selected states 
\begin{eqnarray}
\rho_j&=&p_j^{-1}\,{\cal M}_j^{(1)}\rho{{\cal M}^{(1)}_j}^\dag=\frac{c_j\rho c^\dag_j}{\langle c^\dag_j c_j\rangle}\,, \label{rhk2}\end{eqnarray}
while  $p_j$ is the probability of outcome $j$: 
\begin{eqnarray}
p_j&=&{\rm Tr}\,[\rho\,{{\cal M}^{(1)}_j}^\dag{\cal M}_j^{(1)}]=\langle c^\dag_j c_j\rangle/N\,,\label{pjk2}\end{eqnarray}
with $\sum_{j=1}^D p_j=1$. 
This measurement, with $D$ distinct outcomes, corresponds for instance to the detection of a single fermion through its momentum or sp energy  (labelled by $j$), with $\rho_j$ the ensuing state of remaining fermions. 
Notice that Eq.\ \eqref{maj2} implies 
\begin{equation} 
S_f(\rho^{(M)}_n)\geq S_f[\sum_j p_j \bm{\lambda}(\rho^{(M)}_{jn})]\geq \sum_j p_j S_f(\rho^{(M)}_{jn})\,,\label{ent}
\end{equation}
for any Schur-concave function of $\rho^{(M)}_n$, including  in particular  the  entropies \eqref{Emfn}. Thus, Theorem 1 implies that the entropy $S_f(\rho^{(M)}_n)$ of the normalized $M$-body DMs {\it will not increase, on average, after such operation}. 
For pure states this means that the $M$-body  entropy \eqref{Emfn} will on average not increase, and will in general decrease, after such measurement: 
\begin{equation}
    E_{n_f}^{(M)}(|\Psi\rangle)\geq \sum_j p_j E_{n_f}^{(M)}(|\Psi_j\rangle)\,,
    \label{Efineq}\end{equation}
    where $|\Psi_j\rangle={\cal M}_j|\Psi\rangle/\sqrt{p_j}=c_j|\Psi\rangle/\sqrt{\langle c^\dag_j c_j\rangle}$ is the state after outcome $j$. For mixed states $\rho$ with definite $N$, Eq.\ \eqref{Efineq} implies a similar inequality \begin{equation}
  E^{(M)}_{n_f}(\rho)\geq \sum_j p_j 
  E^{(M)}_{n_f}(\rho_j)\,,\label{Efineq2}\end{equation}
  for the convex-roof extension of \eqref{Emfn},  the $M$-body {\it entanglement of formation} $E^{(M)}_{n_f}(\rho)=
{\rm Min}\sum_i q_i E^{(M)}_{n_f}(|\Psi_i\rangle)$, 
where the minimum is  over all representations \eqref{rhox}  of $\rho$ as convex mixture of pure states with definite  $N$.  Eq.\ \eqref{Efineq2} follows from \eqref{Efineq} by using the representation of $\rho$ minimizing $E_{n_f}^{(M)}(\rho)$, such that 
 $E_{n_f}^{(M)}(\rho)$ $=\sum_i q_i E_{n_f}^{(M)}(|\Psi_i\rangle)\geq 
 \sum_{i,j}q_i p_{ij}E_{n_f}^{(M)}(|\Psi_{ij}\rangle)
 \geq \sum_j p_j E_{n_f}^{(M)}(\rho_j)$, 
 where $p_j=\sum_i q_i p_{ij}$ and $p_{ij}=\langle \Psi_i|c^\dag_j c_j|\Psi_i\rangle/N$. 

{\it Proof of Eq.\ \eqref{maj2}:} Let $\rho$ 
be the state of an $N$ fermion system upon which the operation defined by the operators \eqref{M2} is performed. After outcome $j$ is obtained,  the elements of the $M$-body DM $\rho^{(M)}_{j}$ determined by the ensuing state \eqref{rhk2} are, for $M\leq N-1$, 
\begin{eqnarray}
({\rho^{(M)}_j})_{\bm{\alpha\alpha}'}&=&p_j^{-1}\,{\rm Tr}\,{\cal M}_j^{(1)}
\rho{\cal M}^{(1)\dag}_j C^{(M)\dag}_{\bm\alpha'}C^{(M)}_{\bm\alpha}\label{p1}\\
&=&p_j^{-1}\,{\rm Tr}\,\rho\, C^{(M)\dag}_{\bm\alpha'}{{\cal M}^{(1)}_j}^\dag {\cal M}_j^{(1)} C^{(M)}_{\bm\alpha},\label{p2}
\end{eqnarray}
where the last line holds because operators $C^{(M)}_{\bm\alpha}$ and ${\cal M}_j^{(1)}$ either commute or anticommute. Then 
\begin{equation}
    \sum_j p_j(\rho^{(M)}_j)_{\bm\alpha\bm\alpha'}=
    {\rm Tr}\,[\rho\, C^{(M)\dag}_{\bm\alpha'}
\frac{\hat{N}}{N}C_{\bm\alpha}^{(M)}]=\frac{N-M}{N}\rho^{(M)}_{\bm\alpha\bm\alpha'},\label{eqg}\end{equation}
$\forall$ $\bm\alpha,\bm\alpha'$, implying $\sum_j p_j\rho^{(M)}_j=\frac{N-M}{N}\rho^{(M)}$ 
    and hence 
\begin{equation}
   \rho^{(M)}/{\textstyle\binom{N}{M}}=\sum_j p_j\,\rho^{(M)}_j/{\textstyle\binom{N-1}{M}}\,, 
    \label{eq2}
\end{equation}
for the normalized $M$-body densities, i.e., 
\begin{equation}
    \rho^{(M)}_n=\sum_j p_j \rho^{(M)}_{jn}\,.
\end{equation}
This equation implies the majorization relation \eqref{maj2} 
since \begin{equation}\bm{\lambda}(\sum_i A_i)\prec\sum_i \bm{\lambda}(A_i)\label{majs}\end{equation} for any set of  hermitian matrices $A_i$ of the same dimension  \cite{NV.01,Ni.01,Bh.97,MOA.11}. 
\qed

 In \cite{GDR.20} we have shown that one-body entanglement, i.e., the one quantified by the mixedness of the SPDM 
$\rho^{(1)}$, is also non increasing under 
measurements of the occupancy of a fixed sp state. Such measurement, with just two-possible outcomes ($1$ or $0$),  is  described by the number conserving projection operators 
${\cal P}_j=c^\dag_j c_j\,,\;\;\;{\cal P}_{\bar{j}}=c_jc^\dag_j=\mathbbm{1}-{\cal P}_j$ 
and leads to the majorization relation \cite{GDR.20}
\begin{equation}
    \bm{\lambda}(\rho^{(1)})\prec n_j \bm{\lambda}(\rho^{(1)}_j)+(1-n_j)\bm{\lambda}(\rho^{(1)}_{\bar{j}})\,,\label{maj1}
\end{equation}
between the spectra of the initial and post-selected SPDMs, where $n_j=\langle\Psi|c^\dag_j c_j|\Psi\rangle$  and 
$\rho^{(1)}_j$, $\rho^{(1)}_{\bar{j}}$ are the SPDM's determined by the post-selected states $|\Psi_j\rangle=c^\dag_j c_j|\Psi\rangle/\sqrt{n_j}$ and 
$|\Psi_{\bar j}\rangle=c_jc^\dag_j |\Psi\rangle/\sqrt{1-n_j}$. For an $N$-fermion state $|\Psi\rangle$ an identical relation obviously holds for the normalized DMs $\rho^{(1)}_n=\rho^{(1)}/N$, $\rho^{(1)}_{jn}=\rho^{(1)}_j/N$. 

 However, Eq.\ \eqref{maj1} does not hold, in general, for $\rho^{(M)}$ with $M\geq 2$,  
implying that $M$-body entanglement will not necessarily decrease after such measurement. 
A simple analytic example is provided in  Appendix B. Essentially, measurement of the occupancy of a sp state reduces  the available sp  space for ``collective  pairs'' in a  state like \eqref{ks}, implying a lower maximum eigenvalue $\lambda^{(M)}_1$ in the post-measurement states and hence violation of the inequality \eqref{maj1} by $\rho^{(M)}$ with $M\geq 2$. This result is expected as these measurements do not necessarily increase our knowledge of the $M$-body DM for $M\geq 2$. 

\subsection{$M$-body density operators and generalized majorization relations}
To each $M$-body DM we can associate an $M$-body density operator (DO) 
\begin{equation}
    \hat{\rho}^{(M)}=\sum_{\bm\alpha,\bm\alpha'}\rho_{\bm\alpha\bm\alpha'}^{(M)}C_{\bm\alpha}^{(M)\dag}|0\rangle\langle 0|C_{\bm\alpha'}^{(M)}\,,
    \label{rmo}
\end{equation}
which is the unique mixed state of $M$ fermions satisfying  
\begin{equation} {\rm Tr}\,[\hat{\rho}^{(M)} 
C^{(M)\dag}_{\bm\alpha'}C_{\bm\alpha}^{(M)}]=\rho^{(M)}_{\bm\alpha\bm\alpha'}
\label{rmo2}\end{equation}
$\forall$ $\bm\alpha,\bm\alpha'$, due to  Eq.\ \eqref{nor}. The normal decomposition \eqref{ScDc}--\eqref{diagm} provides its diagonal representation: 
\begin{equation}
    \hat{\rho}^{(M)}=\sum_\nu \lambda_\nu^{(M)}A_\nu^{(M)\dag}|0\rangle\langle 0|A_\nu^{(M)}\,.\label{rmd}\end{equation}

Now consider again the measurement \eqref{M2}  applied on a general mixed state $\rho$ of $N$ fermions.  If the result  is unknown, the post-measurement state (with no post-selection) $\rho'=\sum_j p_j\rho_j$,   is the $N-1$ fermion state 
\begin{equation}\rho'=\sum_j {\cal M}^{(1)}_j\,\rho\, {\cal M}_j^{(1)^\dag}=\frac{1}{N}
\sum_j c_j\,\rho\,c_j^\dag \,.\label{rhop} \end{equation}  
Using Eq.\  \eqref{eqg} for $M=N-1$ we obtain 
\begin{equation}
  {\rm Tr}\,[\rho' \,C^{(N-1)\dag}_{\bm\alpha'}C^{(N-1)}_{\bm\alpha}]=\frac{1}{N}\rho^{(N-1)}_{\bm\alpha\bm\alpha'}\label{rhpp}
\end{equation}
$\forall$ $\bm\alpha,\bm\alpha'$,  implying $\rho'=\hat{\rho}_n^{(N-1)}:=\hat{\rho}^{(N-1)}/N$, the normalized $(N-1)$-body DO 
(${\rm Tr}\,\hat{\rho}^{(N-1)}_n=1$). In summary,   
\begin{equation}
    \hat{\rho}^{(N-1)}=\sum_j c_j \rho c_j^\dag\,,\label{rhn1}
\end{equation}
The $(N-1)$-body DO is then just proportional to the post-measurement state \eqref{rhop}, with the operation in \eqref{rhn1} playing the role of a partial trace.  Expressions similar or equivalent to  \eqref{rhn1} have been previously derived in refs.\ \cite{CK.61,AF.17,SD.19}, with \cite{AF.17}  discussing its extension to states with no fixed fermion number. 

These results, together with Eqs.\ \eqref{maj2}--\eqref{Efineq2}, can be extended to $L$-body measurements,  in which $L$ fermions are  annihilated. Such measurement can be obtained by applying previous measurement $L$ times, i.e., by composing it  with itself $L$ times, and is described by the Kraus  operators 
\begin{equation}
    {\cal M}^{(L)}_{\bm\beta}= \frac{C^{(L)}_{\bm\beta}}{\sqrt{\binom{N}{L}}}= \frac{c_{\beta_1}\ldots c_{\beta_L}}{\sqrt{\binom{N}{L}}}\,,\label{Mkl}
\end{equation}
which, due to Eq.\ \eqref{Nm}  satisfy 
\begin{equation}\sum_{\bm\beta} {\cal M}^{(L)\dag}_{\bm\beta}{\cal M}^{(L)}_{\bm\beta}={\textstyle\frac{1}{\binom{N}{L}}}\sum_{\bm\beta} C^{(L)\dag}_{\bm\beta} C_{\bm\beta}^{(L)}=\mathbbm{1}_N
\end{equation} 
within the subspace of $N$-fermion states. Then, 
\begin{eqnarray}
    {\rm Tr}\sum_{\bm\beta} C_{\bm\beta}^{(L)}\rho\, C_{\bm\beta}^{(L)\dag}C^{(M)\dag}_{\bm\alpha'}C^{(M)}_{\bm\alpha}
  & =&{\textstyle\binom{N-M}{L}}\rho^{(M)}_{\bm\alpha\bm\alpha'} \label{relm}\end{eqnarray}
   $\forall$ $\bm\alpha,\bm\alpha'$ for $M\leq N-L$. Hence, for $L=N-M$ it implies 
   \begin{eqnarray} \hat{\rho}^{(M)}&=&\sum_{\bm\beta} C_{\bm\beta}^{(N-M)}\rho\, C_{\bm\beta}^{(N-M)\dag}\label{rhmg2}\\
   &=&{\textstyle\frac{1}{(N-M)!}}\!\!\!\sum_{j_1,\ldots,j_{N-M}}\!\!\!c_{j_1}\ldots c_{j_{N-M}}\rho\, c^\dag_{j_{N-M}}\ldots c^\dag_{j_1}\nonumber\\
   &=&{\textstyle\binom{N}{N-M}}\sum_{\bm\beta}{\cal M}_{\bm\beta}^{(N-M)}\,
   \rho\,{\cal M}_{\bm\beta}^{(N-M)\dag}\label{rhmg}\,,\end{eqnarray}
   which generalizes Eq.\ \eqref{rhn1}. The sum in \eqref{rhmg} is  the post-measurement state (without post-selection) 
   of the $L=(N-M)$-fermion  measurement \eqref{Mkl}.
   
For general $L\leq N$, this  measurement  has $\binom{D}{L}$ distinct outcomes $\bm\beta$, with probabilities 
\begin{equation}
    p_{\bm\beta}={\rm Tr}\,\rho\,{\cal M}_{\bm\beta}^{(L)\dag}{\cal M}_{\bm\beta}^{(L)}=\rho^{(L)}_{\bm\beta\bm\beta}/{\textstyle\binom{N}{L}}\,,
\end{equation}
satisfying $\sum_{\bm\beta} p_{\bm\beta}=1$, and post-selected  states 
\begin{equation}
    \rho_{\bm\beta}=p_{\bm\beta}^{-1}\,{\cal M}_{\bm\beta}^{(L)}\rho{\cal M}^{(L)\dag}_{\bm\beta}=C_{\bm\beta}^{(L)}\rho\, C_{\bm\beta}^{(L)\dag}/\rho^{(L)}_{\bm\beta\bm\beta}\,,   \end{equation}
    which generalize Eqs.\ \eqref{rhk2}--\eqref{pjk2}. 
From Eq.\ \eqref{relm} we then  obtain, for the ensuing conditional $M$-body DMs of elements  $(\rho^{(M)}_{\bm\beta})_{\bm\alpha\bm\alpha'}={\rm Tr}\,\rho_{\bm\beta}\, C^{(M)\dag}_{\bm\alpha'} C^{(M)}_{\bm\alpha}$, 
\begin{equation}
    \sum_{\bm\beta} p_{\bm\beta} \rho^{(M)}_{\bm\beta}=\frac{\binom{N-M}{L}}{\binom{N}{L}}\rho^{(M)}\,,\label{rmx}
\end{equation}
for $M\leq N-L$. Eq.\ \eqref{rmx} is equivalent to 
\begin{equation}
    \sum_{\bm\beta} p_{\bm\beta} \rho^{(M)}_{\bm\beta}/{\textstyle\binom{N-L}{M}}=\rho^{(M)}/{\textstyle\binom{N}{M}}\,,\label{rmx2}
\end{equation}
and hence to 
\begin{equation}
    \rho^{(M)}_n=\sum_\beta p_{\bm\beta} \rho^{(M)}_{\bm\beta n}\,,\label{rmx3}
\end{equation}
for the normalized $M$-body DMs. Eq. \eqref{rmx3} then implies the general majorization relation
\begin{equation} \bm{\lambda}(\rho^{(M)}_n)\prec \sum_\beta p_{\bm\beta} 
\bm{\lambda}(\rho^{(M)}_{\bm\beta n})\label{mgral}\,,\end{equation}
between the initial and post-measurement conditional normalized $M$-body DMs, which generalizes Eq.\  \eqref{maj2} to the measurement described by the operators \eqref{Mkl}. 
The associated entropies satisfy 
\begin{equation}
   S_{f}(\rho^{(M)}_n)\geq \sum_{\bm\beta} p_{\bm\beta} S_f(\rho^{(M)}_{\bm\beta n})\,.
     \label{sfgral}\end{equation} 
   For pure states this implies that the  entanglement entropies \eqref{Emfn} will not increase, on average, after these measurements. 
   The same occurs with the associated entanglement of formation for initial mixed states. 

\subsection{Mapping to bipartite systems}

In the previous sections it was proposed to link the mixedness of either $\rho^{(M)}$ or $\rho^{(N-M)}$, as reduced DM's of a given $N$-fermion state, to the amount of correlations between $M$-body and $N-M$-body observables on such state. In this  section we will show this relation is operationally justified by the existence of a class of quantum maps converting states of indistinguishable fermions into states of effectively distinguishable fermions, in such a way that entanglement in the target state is bounded by  the entropy \eqref{Emfn} of the normalized $M$-body  DM. 

Let $|\Psi\rangle$ be an $N$-fermion state with support in a sp subspace ${\cal H}$ of dimension $D\geq N$ and let ${\cal H}_A$ be  a sp subspace of dimension $D_A\geq M$ {\it orthogonal} to ${\cal H}$, such that $\{c_{i_A},c_j\}=\{c^\dag_{i_A},c^\dag_j\}=\{c_{i_A},c^{\dag}_j\}=0$ for $i_A$ ($j$) labelling sp states in ${\cal H}_A$ (${\cal H}$) (entailing  $\langle i_A|j\rangle=\langle 0|c_{i_A}c^\dag_j|0\rangle=0$ $\forall\, i_A,j$). They are of course subspaces of a complete sp space ${\cal H}_T$ such that ${\cal H}\oplus {\cal H}_A\subset{\cal H}_T$.    
 Consider now a CPTP map ${\cal T}_M$ described by Kraus operators 
\begin{eqnarray}
{\cal T}^r&=&{\textstyle\frac{1}{\sqrt{\binom{N}{M}}}}\sum_{\bm\mu,\bm\alpha} T^r_{\bm\mu\bm\alpha}\, C^{(M)\dag}_{\bm\mu} C^{(M)}_{\bm\alpha}\,,\label{Tr}
\end{eqnarray}
where $C^{(M)\dag}_{\bm\mu}=(c^\dag_{i_{1_A}},\ldots, c^\dag_{i_{M_A}})$  creates $M$ fermions in  ${\cal H}_A$ while $C^{(M)}_{\bm\alpha}=(c_{j_M},\ldots,c_{j_1})$  annihilates $M$ fermions in ${\cal H}$, and $T^r$ is a $\binom{D_A}{M}\times\binom{D}{M}$  matrix.   Assuming ${\cal H}_A$ initially ``empty'' and the condition 
$\sum_r T^{r\dag}T^r=\mathbbm{1}$,  they will satisfy 
\begin{equation*}
    \sum_r {\cal T}^{r\dag}{\cal T}^r={\textstyle\frac{1}{\binom{N}{M}}}
    \sum_{\bm\alpha,\bm\alpha',r}(T^{r\dag}T^r)_{\bm\alpha'\bm\alpha}C^{(M)\dag}_{\bm\alpha'}C^{(M)}_{\bm\alpha}=\mathbbm{1}_N
\end{equation*} 
by virtue of Eq.\ \eqref{Nm}, within 
the Fock space ${\cal F}_N({\cal H})$ of $N$ fermion states with support in ${\cal H}$. Its action on an $N$-fermion state $|\Psi\rangle\in{\cal F}_N({\cal H})$ is, using Eqs.\ \eqref{Psm} and \eqref{3}, 
\begin{eqnarray}
    {\cal T}^r|\Psi\rangle&=&{\textstyle\frac{1}{\sqrt{\binom{N}{M}}}}
    \sum_{\bm\mu,\bm\alpha,\bm\beta} T^r_{\bm\mu\bm\alpha}\Gamma^{(M)}_{\bm\alpha\bm\beta} C_{\bm\mu}^{(M)\dag}C_{\bm\beta}^{(N-M)\dag}|0\rangle\nonumber\\
    &=&\sum_{\bm\mu,\bm\beta}\Gamma^{r}_{\bm\mu\bm\beta}C_{\bm\mu}^{(M)\dag}C_{\bm\beta}^{(N-M)\dag}|0\rangle\,,\label{KPsi}
\end{eqnarray}
where $\Gamma^r=T^r\Gamma^{(M)}/\sqrt{\binom{N}{M}}$. It is verified that the probability  of outcome $r$, 
\begin{equation} p_r=\langle\Psi| {\cal T}^{r\dag}{\cal T}^r|\Psi\rangle={\rm Tr}(\Gamma^{r\dag}\Gamma^{r})\label{pr}\,,
\end{equation} 
satisfies $\sum_{r} p_{r}= {\frac{1}{\binom{N}{M}}}{\rm Tr}(\Gamma^{(M)\dag}\Gamma^{(M)})=1$ due to Eq.\ \eqref{trm}. 

Operators \eqref{Tr} can therefore be regarded as Kraus operators describing a CPTP map on the $N$-fermion Fock space that takes, with probability $p_{r}$, the original state $|\Psi\rangle$ into the $(M, N-M)$ ``bipartite'' state $|\Phi_{AB}^{r}\rangle=(\sqrt{p_{r}})^{-1}{\cal T}^{r}|\Psi\rangle$  containing  $N-M$ fermions in ${\cal H}$ 
and $M$ fermions in the orthogonal sp space ${\cal H}_A$. 
 
After the map is implemented with outcome $r$, we can look at the reduced state  of the fermionic modes in ${\cal H}_A$  in the state  $|\Phi_{AB}^{r}\rangle$ (which will contain $M$ fermions), 
${\rho}_A^{r}={\rm Tr}_{B} |\Phi^{r}_{AB}\rangle\langle\Phi^{r}_{AB}|$, 
such that $\langle \Phi_{AB}^{r}|O_A|\Phi_{AB}^{r}\rangle={\rm Tr}\,[\rho_A^r O_A]$ for any number conserving  operator $O_A$  depending just on operators $c^\dag_{i_A}$, $c_{i_A}$ with support in ${\cal H}_A$. Using \eqref{KPsi} we obtain  ${\rho}_A^{r}=\sum_{\bm\mu,\bm\mu'}(\rho_A^{r})_{\bm\mu\bm\mu'}C^{(M)\dag}_{\bm\mu}|0\rangle\langle 0|C^{(M)}_{\bm\mu'}$ with 
\begin{equation}
    \rho_A^{r}=
    \Gamma^{r}\Gamma^{r\dag}/p_{r}\,.\label{Mrars}
\end{equation}
Similarly $\rho_B^{r}=
    \Gamma^{rT}\Gamma^{r*}/p_{r}$. 

As a direct consequence of expression  \eqref{Mrars} we can derive the following important result: If $\bm\lambda(\rho^{r}_A)$ and $\bm\lambda(\rho^{(M)}_n)$ denote again the eigenvalue vectors (sorted in decreasing order) of $\rho_A^{r}$ and $\rho^{(M)}_n=\rho^{(M)}/\binom{N}{M}$ respectively, then the following majorization relation holds:
\begin{equation}
    \bm\lambda(\rho^{(M)}_n)\prec\sum_{r} p_{r}\bm\lambda(\rho_A^{r}),\label{majave}
\end{equation}
where the vector of smaller dimension is assumed to be completed with zeros to match the dimensions. 
\begin{proof}
From  Eqs.\ \eqref{Mrars}  and \eqref{mDM} we obtain 
\begin{equation*}
    p_r\rho_A^r={\textstyle\frac{1}{\binom{N}{M}}} T^r \Gamma^{(M)}\Gamma^{(M)\dag} T^{r\dag}= T^r \rho^{(M)}_n T^{r\dag}\,. 
\end{equation*}
We can also write 
\begin{equation*} \rho^{(M)}_n=\sum_r\sqrt{\rho^{(M)}_n}T^{r\dag}T^r\sqrt{\rho^{(M)}_n}\label{romn2}\,,\end{equation*}
which implies, using again \eqref{majs}, 
\begin{equation}
    \bm{\lambda}(\rho^{(M)}_n)\prec\sum_r \bm{\lambda}\left(\sqrt{\rho^{(M)}_n}T^{r\dag}T^r\sqrt{\rho^{(M)}_n}\right)=\sum_r p_r\bm{\lambda}(\rho_A^r)\label{Mm2}
\end{equation}
since the nonzero eigenvalues of 
$(\sqrt{\rho^{(M)}_n}T^{r\dag})(T^r\sqrt{\rho^{(M)}_n})$ are the same as those of 
$T^r\sqrt{\rho^{(M)}_n}\sqrt{\rho^{(M)}_n}T^{r\dag}=p_r\rho_{A}^r$.  
\end{proof}

In particular, Eq.\ \eqref{majave} implies that the average bipartite  entanglement entropy between $A$ and $B$ of the final states, defined  by $\sum_{r} p_{r}E_f(|\Phi^{r}_{AB}\rangle)$ 
with $E_f(|\Phi^{r}_{AB}\rangle=S_f(\rho_A^{r})=S_f(\rho_B^{r})$,  {\it will never exceed the $M$-body  entropy \eqref{Emfn}}, which provides,  therefore, an {\it upper bound}  to the generated bipartite entanglement: 
\begin{equation}
    E_{f_n}^{(M)}(|\Psi\rangle)=S_f(\rho_n^{(M)})\geq \sum_{r} p_{r}S_f(\rho_A^{r}) \,.
\end{equation}
A second immediate corollary of \eqref{majave} is that  we can derive a condition for the conversion of a pure $N$-fermion state $|\Psi\rangle$ into a bipartite state $|\Phi_{AB}\rangle$ of $M$ and $N-M$ fermions in orthogonal sp spaces, i.e., with $M$ fermions occupying sp states in ${\cal H}_A$, and $N-M$ in ${\cal H}$,  by means of a CPTP map described by the  operators \eqref{Tr}: for such conversion to be possible the relation
\begin{equation}
    \bm\lambda(\rho^{(M)}_n)\prec\bm\lambda(\rho_A),\label{prec2}
\end{equation}
must hold, where $\rho_A$ is the reduced state of $A$ in $|\Phi_{AB}\rangle$.
\begin{proof}
If the map takes $|\Psi\rangle$ into $|\Phi_{AB}\rangle$ and  is described by the operators \eqref{Tr}, then for all $r$ we have
\begin{equation}
    {\cal T}^{r}|\Psi\rangle=\sqrt{p_{r}}|\Phi_{AB}\rangle,
\end{equation}
with $p_{r}$ given by \eqref{pr}. It follows  that $\rho^{r}_A=\rho_A$ $\forall$ $r$, and by  virtue of \eqref{majave}, Eq.\ \eqref{prec2} follows.  
\end{proof}

We have therefore identified a class of quantum maps that could be used to transform a pure state $|\Psi\rangle$ of $N$ indistinguishable fermions into another state $|\Phi_{AB}\rangle$ which contains a definite number of fermions in two orthogonal sp subspaces, such that 
$M$ fermions can be effectively ``distinguished'' from the remaining $N-M$ due the orthogonality (for instance, $M$ fermions confined in a spatial  region well separated from that of the remaining $N-M$ fermions, or in orbitals orthogonal to those occupied in $|\Psi\rangle$).  The conversion is such that the entanglement between these two sets of particles in the target state is bounded from above, on average, by the entropy of the normalized $M$-body DM. This provides  a clear operational meaning to the suggested link between the mixedness of the eigenvalues of this matrix and the amount of correlations between the indistinguishable particles in $|\Psi\rangle$.

As a trivial example, consider  a two fermion state 
\begin{equation}|\Psi\rangle=(\alpha c^\dag_1c^\dag_2+\beta c^\dag_3 c^\dag_4)|0\rangle\,.\end{equation}
As mentioned in \ref{IIB}, any two-fermion state with support in a sp subspace of dimension $4$ (i.e.\ occupying just 4 sp levels) can be written in this way \cite{SL.01,SC.01,GR.15}. 
Its SPDM $\rho^{(1)}$ is diagonal in this basis, with $\langle c^\dag_{j}c_i\rangle=\delta_{ij}f_i$ and $f_i=|\alpha|^2$ for $i=1,2$, $f_i=|\beta^2|$ for $i=3,4$ and $f_i=0$ otherwise,  such that its nonzero eigenvalues are  $\bm{\lambda}(\rho^{(1)})=(|\alpha|^2,|\alpha|^2,
|\beta|^2,|\beta|^2)$, with $|\alpha|^2+|\beta|^2=1$.  
Consider now the simple map ${\cal T}=\frac{1}{\sqrt{2}}\sum_i c^\dag_{i_A}c_{i}$, where $c_i$ annihilates a fermion in ${\cal H}$ ($i=1,\ldots,D$) while 
$c^\dag_{i_{A}}$  creates a fermion  in ${\cal H}_A$, initially empty (and of dimension $D_A\geq D$). Then ${\cal T}^\dag{\cal T}=\mathbbm{1}_2$ in ${\cal F}_2({\cal H})$ and  \begin{equation}
   {\cal T}|\Psi\rangle=\frac{1}{\sqrt{2}}[\alpha(c^\dag_{1_A} c^\dag_{2}-c^\dag_{2_A}c^\dag_{1})+\beta(c^\dag_{3_A} c^\dag_{4}-c^\dag_{4_A}c^\dag_{3})]|0\rangle \end{equation} 
 is a normalized two-fermion state with one fermion in ${\cal H}_A$ and one in ${\cal H}$. It leads  to reduced states $\rho_{A(B)}$ with spectrum $\bm{\lambda}(\rho_{A(B)})=(|\alpha|^2,|\alpha|^2,|\beta|^2,|\beta|^2)/2$ identical to that of $\rho^{(1)}/2$, thus saturating the inequalities of the majorization relation \eqref{majave}. We note that even if $\beta=0$, in which case $|\Psi\rangle$ is a SD, ${\cal T}|\Psi\rangle$ is no longer a SD and has therefore non-zero one-body entanglement, in agreement with the result that bipartite entanglement with fixed fermion number or number parity at each side requires one-body entanglement \cite{GDR.20}. 

In contrast, for  ${\cal T}^r=\frac{1}{\sqrt{2}}c^\dag_{1_A} c_r$, 
 $r=1,\ldots D$, 
still $\sum_r {\cal T}^{r\dag}{\cal T}^r=\mathbbm{1}_2$ in ${\cal F}_2({\cal H})$ but 
${\cal T}^{r}|\Psi\rangle$ is either $0$ or proportional to a state $c^\dag_{1_A}c^\dag_{j}|0\rangle$ with $c^\dag_j|0\rangle\in{\cal H}$,  and no direct $A-B$ entanglement is generated, thus trivially satisfying Eq.\ \eqref{majave}. 

\section{Conclusions}
We have examined  a general bipartite-like  representation and  Schmidt decomposition of arbitrary $N$-fermion states, which expresses it in terms of general $M$- and ($N$-$M$)-fermion creation operators. It is directly connected to the reduced $M$ and ($N$-$M$)-body DMs, which share the same non-zero eigenvalues,  formally resembling  the standard case of distinguishable components. It naturally leads to the concept of $M$-body entanglement, determined by the mixedness of the $M$ or ($N$-$M$)-body DMs,  which generalizes one-body entanglement and characterizes the correlations  between $M$- and $(N$-$M)$-body operators. Such entanglement is of course independent of the choice of sp basis (it is not a mode-entanglement) and in fact also of the choice of $M$-fermion creation operators, as long as they satisfy Eqs.\ \eqref{nor2}--\eqref{Nm2}. 

The full set of $M$-body DM's $\rho^{(M)}$ ($1\leq M\leq N/2$) provides a detailed characterization  of the structure of correlated $N$-fermion states. We have explicitly seen that correlated states which look similar at the level of the one-body DM,  
can lead to very distinct $M$-body DMs, as in the case of the states  \eqref{GHZ} and \eqref{ks}. In the latter,  emergence of bosonic-like features is signaled by the appearance of an eigenvalue larger than $1$ in the $M$-body DM for  $M\geq 2$.  On the other hand, all SDs  (i.e.\ all free- or independent-fermion states) lead to idempotent $M$-body DMs (i.e.\ with eigenvalues $1$ or $0$) $\forall$ $M$. The eigenvalues of $\rho^{(M)}$  determine the average of $M$-body operators  and its largest eigenvalue provides an upper bound to all averages $\langle A^{(M)\dag} A^{(M)}\rangle$ (Eq.\ \eqref{bnd}) determined by  ``collective'' operators $A^{(M)\dag}$ creating a normalized state of $M$ fermions. 

Finally, we have investigated some  operational implications of $M$-body entanglement. By demonstrating the majorization relation \eqref{maj2}, we have shown that  the entropy \eqref{Emfn} of the normalized $M$-body DM  
will not increase (and will in general decrease)  under single fermion  measurements determined by the Kraus operators \eqref{M2}. This result can be extended to general $L$-fermion measurements based on the operators \eqref{Mkl} (Eqs.\ \eqref{mgral}--\eqref{sfgral}), 
which  have the reduced $M$-body DMs  as  post-measurement states.  Moreover, 
by proving the majorization relation \eqref{majave}, we have shown that that such $M$-body entropy also provides an upper bound to the average bipartite entanglement  entropy between $M$ and $N-M$ effectively distinguishable  fermions generated by the mapping determined by the operators \eqref{Tr}. Present results then provide the basis for a general theory of  many-body entanglement beyond anti-symmetrization in fermion systems.

\acknowledgments
The authors acknowledge support from CONICET (NG, MDT) and CIC (RR) of Argentina. Work
supported by CONICET PIP Grant No. 11220150100732.

\appendix
\section{Eigenvalues  of $\rho^{(M)}$  in the states $|\Psi_{2k}\rangle$}
We first derive here the eigenvalues of the first three $M$-body DM's in the states \eqref{ks}.  Since fermions are created in pairs $c^\dag_{2i-1}c^\dag_{2i}$ with equal probability, the elements of the one-body DM  in these states are easily seen to be 
\begin{equation} \langle\Psi_{2k}|c^\dag_{i}c_j|\Psi_{2k}\rangle=\delta_{ij}N/D\,,\end{equation}
implying $\rho^{(1)}=\frac{N}{D}\mathbbm{1}$.  Thus, it is the maximally mixed SPDM compatible with the total fermion number $N$, being hence diagonal in {\it any} sp basis with a single $D$-fold degenerate eigenvalue $N/D$. 

On the other hand, the elements  of the two-body DM are blocked in two submatrices. The first one, comprising the contiguous pair creation operators $c^\dag_{2i-1}c^\dag_{2i}$ that form  the operator $A^\dag$ of Eq.\ \eqref{Ad}, has elements 
\begin{equation}\langle\Psi_{2k}| c^\dag_{2i-1}c^\dag_{2i}c_{2j}c_{2j-1}|\Psi_{2k}\rangle=
\alpha\delta_{ij}+\beta(1-\delta_{ij})\,,\end{equation}
where, using  $k=N/2$ and assuming $D$ even, 
\[\alpha=
{\textstyle\frac{\binom{D/2-1}{k-1}}{\binom{D/2}{k}}=\frac{2k}{D}\,,
\;\;\beta=\frac{\binom{D/2-2}{k-1}}{\binom{D/2}{k}}=
\frac{2k(D-2k)}{D(D-2)}\,.}\]
Hence, since this $\frac{D}{2}\times \frac{D}{2}$ block is just 
 $(\alpha-\beta)\mathbbm{1}+M$, with $M$ a rank one matrix with all elements equal to $\beta$, it has just two distinct eigenvalues: a   nondegenerate eigenvalue 
\begin{equation}\lambda^{(2)}_{\rm max}=\alpha+(\frac{D}{2}-1)\beta=k(1-2(k-1)/D)\,,\label{la12a}\end{equation}
which is precisely that associated with the collective uniformly weighted pair creation operator $A^\dag$: 
\begin{equation} \langle\Psi_{2k}|A^\dag A|\Psi_{2k}\rangle=\lambda^{(2)}_{\rm max}\,,
\end{equation}
and a $(D/2-1)$-fold degenerate smaller eigenvalue  
\begin{equation}\lambda^{(2)}_{\rm rest}=\alpha-\beta=\frac{4k(k-1)}{D(D-2)}\,.\end{equation}
The other block comprises the remaining $\binom{D}{2}-\frac{D}{2}$ pair creation operators $c^{\dag}_i c^\dag_j$ involving two distinct pairs and is directly diagonal, 
with elements $\frac{\binom{D/2-2}{k-2}}{\binom{D/2}{k}}=\lambda^{(2)}_{\rm rest}$. Thus, the final result is one large nondegenerate eigenvalue $\lambda^{(2)}_{\rm max}\geq 1$,   plus $\binom{D}{2}-1$ smaller identical eigenvalues $\lambda^{(2)}_{\rm rest}<1$, satisfying 
\begin{equation} {\textstyle\lambda^{(2)}_{\rm max}+\left(\binom{D}{2}-1\right)\lambda^{(2)}_{\rm rest}=\binom{N}{2}\,.}\end{equation}

The same procedure can be applied to determine the eigenvalues of the three-body DM $\rho^{(3)}$. 
For creation of three fermions with two of them in one of the contiguous pairs ($2i-1,2i$),  we obtain 
\begin{equation}\langle\Psi_{2k}| c^\dag_{2i-1}c^\dag_{2i}c^\dag_{j}c_{k} c_{2l}c_{2l-1}|\Psi_{2k}\rangle=\delta_{jk}
[\gamma\delta_{il}+\eta(1-\delta_{il})]\end{equation}
where $k\neq 2l$, $k\neq 2l-1$,  $j\neq 2i$, $j\neq 2i-1$, and 
\[\gamma=
\frac{\binom{D/2-2}{k-2}}{\binom{D/2}{k}}\,,\;\;\;\;
\eta=\frac{\binom{D/2-3}{k-2}}{\binom{D/2}{k}}\,.\]
Hence, we obtain $D$ identical $(\frac{D}{2}-1)\times(\frac{D}{2}-1)$ blocks, each of which  has a large non-degenerate eigenvalue \begin{eqnarray}\lambda^{(3)}_{\rm max}&=&{\textstyle\gamma+(\frac{D}{2}-2)\eta=
\frac{2k(k-1)(1-2(k-1)/D)}{D-2}}\label{A8}\;\;\;\;\;\;\;\\
&=&\langle\Psi_{2k}|A^{(3)\dag}_jA^{(3)}_j|\Psi_{2k}\rangle,\end{eqnarray}
where  $A^{(3)\dag}_j=\frac{1}{\sqrt{D/2-1}}\sum_{i}c^\dag_{2i-1}c^\dag_{2i}c^\dag_j$, 
  and $D/2-2$ smaller identical eigenvalues
\begin{equation}\lambda^{(3)}_{\rm rest}=
\gamma-\eta={\textstyle\frac{8k(k-1)(k-2)}{D(D-2)(D-4)}}\,,\end{equation}
associated to orthogonal operators 
$A^{(3)\dag}_\nu$. On the other hand, remaining $\binom{D}{3}-D(\frac{D}{2}-1)$ triplets $c^\dag_i c^\dag_j c^\dag_k$ belonging to distinct pairs lead to a diagonal block in $\rho^{(3)}$ with identical diagonal elements $\frac{\binom{D/2-3}{k-3}}{\binom{D/2}{k}}=\lambda^{(3)}_{\rm rest}$. Therefore, there are $D$ eigenvalues equal to $\lambda^{(3)}_{\rm max}$ plus $\binom{D}{3}-D$ eigenvalues equal to $\lambda^{(3)}_{\rm rest}$, satisfying 
\begin{equation} {\textstyle D\lambda^{(3)}_{\rm max}+\left(\binom{D}{3}-D\right)\lambda^{(2)}_{\rm rest}=\binom{N}{3}\,.}\end{equation}
It should be noticed that while $\lambda_{\rm rest}^{(3)}\leq 1$, 
$\lambda_{\rm max}^{(3)}\geq 1$ for $1+\sqrt{D/2}\leq k\leq D/2$, 
reaching its maximum for $k\approx D/3$ for $D\geq 6$,  where 
$\lambda^{(3)}_{\rm max}\approx 2D/27$.

For obtaining the largest eigenvalue $\lambda^{(2m)}_{\rm max}$ 
of the $M=2m$-body DM algebraically, we may directly note that it will arise from a   $\binom{D/2}{m}\times\binom{D/2}{m}$ block containing products of $m$ distinct pairs $c^\dag_{2i-1}c^\dag_{2i}$ and 
$c_{2j}c_{2j-1}$. All elements 
$\langle \Psi_{2k}|c^\dag_{2i_1-1}c^\dag_{2i_1}\ldots c_{2j_1}c_{2j_1-1} |\Psi_{2k}\rangle$ will be positive with all rows of this block having the same sum owing to symmetry. Its largest eigenvalue will then be equal to this sum (as verified in \eqref{la12a} for $m=1$) and is associated to the uniform eigenvector $\propto(1,1,\ldots,1)$, i.e.\ to the collective operator $A^{(2m)\dag}\propto (A^\dag)^m$,  of Eq.\ \eqref{ks}, implying Eq.\ \eqref{l2m}. 
\section{Behavior of  $\rho^{(2)}$ under single mode occupancy measurement}
We will now prove that in the states \eqref{ks}, 
measurement of the occupancy of one sp mode through the operators ${\cal P}_k=c^\dag_k c_k$ and ${\cal P}_{\bar{k}}=c_kc^\dag_k$,  will break the analogous of Eq.\  \eqref{maj1} for $\rho^{(2)}$.  We will prove in fact that the largest eigenvalue \eqref{la12a} of $\rho^{(2)}$ is greater than  that of $\rho^{(2)}_k$ and $\rho^{(2)}_{\bar{k}}$, implying 
\begin{equation}\lambda^{(2)}_1> p_k\lambda^{(2)}_{1k}+(1-p_k)\lambda^{(2)}_{1\bar{k}}\label{ineqk}\,,
\end{equation}
which breaks the first majorization inequality. \\
{\it Proof:} If sp state $k$ is measured to be occupied, which will occur with probability $p_k=N/D$, the associated contiguous pair $(k,k+1)$ ($k$ odd) or $(k-1,k)$ ($k$ even)  becomes ``freezed'' and 
the maximum eigenvalue $\lambda^{(2)}_{k\,{\rm max}}$ of the ensuing DM $\rho^{(2)}_k$ will 
arise from the remaining $N-2$ fermions occupying the other  $D-2$ sp states. Consequently, using Eq.\ \eqref{la12a} for $D\rightarrow D-2$ and $k=N/2\rightarrow N/2-1$, 
\[{\textstyle\lambda^{(2)}_{k\,{\rm max}}=\frac{N-2}{2(D-2)}(D+2-N)<\frac{N}{2D}(D+2-N)=\lambda^{(2)}_{\rm max}}\]
where the inequality holds for $N<D$. 
Similarly, if state $k$ is found to be empty, a similar reasoning leads to 
\[\lambda^{(2)}_{\bar{k}\,{\rm max}}=\frac{N}{2(D-2)}(D-N)<\frac{N}{2D}(D+2-N)=\lambda^{(2)}_{\rm max}\]
where the inequality holds for $N>2$. These two results imply Eq.\ \eqref{ineqk} and hence violation of the majorization relation similar to  \eqref{maj1} for $M=2$. Analogous results can be obtained for $M=3$ in the same states \eqref{ks}. 

For completeness, we also verify that for the measurement \eqref{M2} in the same states \eqref{ks}, 
the first inequality 
\begin{equation}\frac{\lambda_{\rm max}^{(2)}}{\binom{N}{2}}\leq \sum_k p_k \frac{\lambda^{(2)}_{k\,{\rm max}}}{\binom{N-1}{2}}\,,\label{apc}\end{equation}
in \eqref{maj2} for the largest eigenvalues of the initial and post-measurement normalized two-body DMs,   does hold: 
Using again Eq.\ \eqref{la12a} for $\lambda_{\rm max}^{(2)}$ (with $k=N/2$) and $\lambda_{k\,{\rm max}}^{(2)}$ (with $k=N/2-1$ and $D\rightarrow D-2$), with $p_k=\frac{\langle c^\dag_k c_k\rangle}{N}=\frac{1}{D}$, we obtain, in agreement with \eqref{apc},
\[\tfrac{1}{D}\sum_k \tfrac{(N-2)(D+2-N)}{2(D-2)\binom{N-1}{2}}=\tfrac{D+2-N}{(D-2)(N-1)}\geq \tfrac{(D+2-N)}{D(N-1)}=\tfrac{\lambda_{\rm max}^{(2)}}{\binom{N}{2}}.\]

%\bibliographystyle{apsrev4-1}
%\bibliography{bibtex.bib}
%\end{document}

%merlin.mbs apsrev4-1.bst 2010-07-25 4.21a (PWD, AO, DPC) hacked
%Control: key (0)
%Control: author (0) dotless jnrlst
%Control: editor formatted (1) identically to author
%Control: production of article title (0) allowed
%Control: page (1) range
%Control: year (0) verbatim
%Control: production of eprint (0) enabled
%
\end{document}